\documentclass{birkmult}
 \theoremstyle{definition}
 
 \theoremstyle{remark}

 \numberwithin{equation}{section}

 \newcommand{\gvec}[1]{\ensuremath{\mbox{\textbf{\textit{#1}}}}}
 \newcommand{\topline}{\hline}
 \newcommand{\midline}{\hline}
 \newcommand{\bottomline}{\hline}

 \renewcommand{\oe}{\"{o}{}}

 \usepackage{graphicx}
 \usepackage{ifpdf}
 \usepackage[noadjust]{cite}
 \usepackage{amsmath}
 \usepackage{amsfonts}

\begin{document}

\DeclareGraphicsExtensions{.pdf}

%-------------------------------------------------------------------------
% editorial commands: to be inserted by the editorial office
%
%\firstpage{1}
%\volume{228}
%\DOI{003-0001}
%\seriesextraline{This is the Concrete Title of this Book\br H.E. R and S.T.C. W, Eds.}
%\firstpage{1}
\title[New views of crystal symmetry -- Grassmann and Clifford]
 {New views of crystal symmetry guided by profound admiration of the extraordinary works of Grassmann and Clifford}
%----------Author 1
\author[E. Hitzer]{Eckhard Hitzer}

\address{%
Department of Applied Physics\br
University of Fukui\br
910-8507 Fukui\br
Japan
}

\email{hitzer@mech.u-fukui.ac.jp}

\thanks{SDG}


\date{November 27, 2009}
\dedicatory{Soli Deo Gloria}
%%% ----------------------------------------------------------------------

\begin{abstract}
This paper shows how beginning with Justus Grassmann's work, Hermann Grassmann was influenced in his mathematical thinking by crystallography. H. Grassmann's Ausdehnungslehre in turn had a decisive influence on W.K. Clifford in the genesis of geometric algebras. Geometric algebras have been expanded to conformal geometric algebras, which provide an ideal framework for modern computer graphics. Within this framework a new visualization of three-dimensional crystallographic space groups has been created. The complex beauty of this new visualization is shown by a range of images of a diamond cell. Mathematical details are given in an appendix.
\end{abstract}

%%% ----------------------------------------------------------------------
\maketitle
%%% ----------------------------------------------------------------------
%\tableofcontents

\section{Introduction}

Already Hermann Grassmann's father Justus (1829, 1830) published two works on the geometrical description of crystals, influenced by the earlier works of C.S. Weiss (1780-1856) on three main crystal forces governing crystal formation. In his 1840 essay on the derivation of crystal shapes from the general law of crystal formation Hermann established the notion of a three-dimensional vectorial system of forces with rational coefficients, that represent the interior crystal structure, regulate its formation, its shape and physical behavior. In the Ausdehnungslehre 
1844 (\S 171)\cite{HG:AL1844} 
he finally writes: 
\textit{I shall conclude this presentation by one of the most beautiful applications which can be made of the science treated, i.e. the application to crystal 
figures} \cite{ES:JGcrystwork}. 
The geometry of crystals thus certainly influenced the Ausdehnungslehre. 

Grassmann's work in turn influenced 
W.K. Clifford \cite{WC:AppGExtAlg}
in England: 
\textit{I propose to communicate in a brief form some applications of Grassmann's theory \ldots I may, perhaps, therefore be permitted to express my profound admiration of that extraordinary work, and my conviction that its principles will exercise a vast influence upon the future of mathematical science.} 
Conformal Clifford (geometric) algebra has in turn led at the beginning of the 20th century to a new fully geometric description of crystal symmetry in terms of socalled 
versors \cite{HH:CSG}. 
Versors are simply (Clifford) geometric products of five-dimensional vectors conformally representing general planes in three-dimensional (3D) Euclidean space (by their 3D normal vector and the directed distance from the origin). Each plane's vector geometrically represents a reflection at the plane, the geometric products of several plane vectors represents the combination of reflections at the respective planes (Cartan-Dieudonn$\acute{\mathrm{e}}$). 

As expected three crystal specific 3D vectors are enough to construct all symmetry versors of any type of crystal. With the geometric algebra capable graphics software CLUCalc~\cite{CP:CLUCalc} this concept can be implemented in every detail, such that the abstract beauty of the enormously rich symmetry of crystals can be fully visualized by state-of-the-art 3D computer graphics: The Space Group Visualizer (SGV), a tailor-made 
CLUCalc Script \cite{HP:3vecgen,HP:ICCA8,HP:SGV,HP:cellat,HP:crystGA,HP:intvis,HP:sym3vec}. 
To be precise, the SGV is thus capable of showing every plane of reflection and glide-reflection symmetry, all axis of rotations, screw-rotations and rotary inversions, and every center of inversion. It further allows to dynamically visualize the action of any symmetry operation on a general element (representing atoms, molecules or ions). 

We thus have, 165 years after the Ausdehnungslehre of 1844, an explicit form of the beauty, which Grassmann may have had in mind, when he wrote eloquently: 
\textit{one of the most beautiful applications.}

The next section uses the symmetries of diamond in order to demonstrate how the SGV visualizes space group symmetry. For mathematically interested readers the appendix introduces the Clifford geometric algebra description of crystallographic space groups.

\section{Computer visualization of crystal symmetry}

Geometrically a diamond cell lattice (type: face centered cubic = fcc) is highly symmetric. That means there is an enormous variety of possible geometric transformations, that leave the lattice as a whole invariant, including all lengths and angles. These symmetry operations include single cell transformations that leave a cell vertex point invariant: planes of reflections (through the vertex), rotations (with axis trough the vertex, and inversions ($\gvec{x} \mapsto -\gvec{x}$, centered at the vertex), and rotoinversions (inversions followed by a rotation). The 24 symmetry transformations of a diamond vertex point group create 24 symmetric copies of a general asymmetric element placed next to the invariant point, 
\begin{figure}
\begin{center}
  \resizebox{0.5\textwidth}{!}{\includegraphics{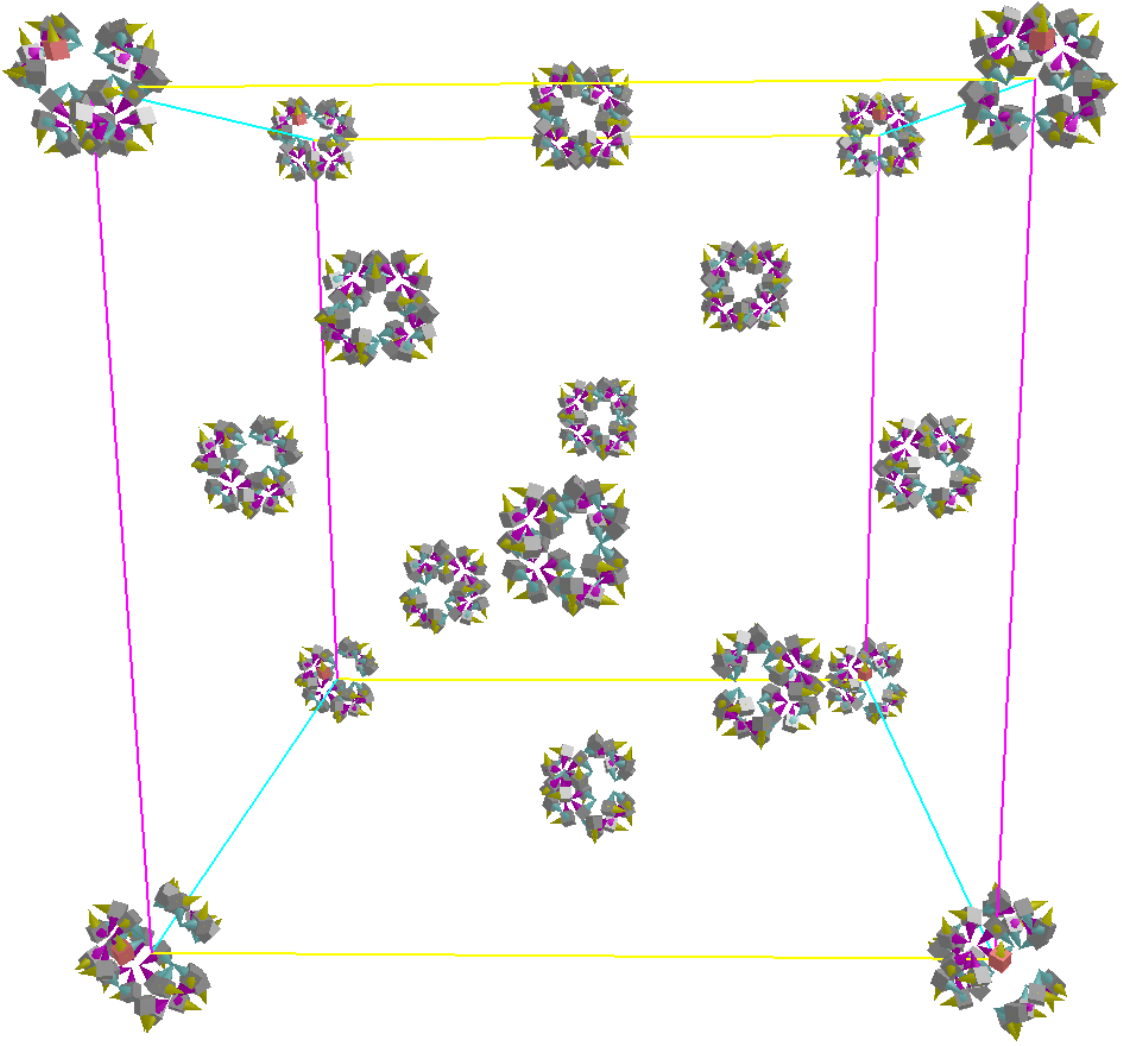}}
  \hspace{2mm}
  \resizebox{0.45\textwidth}{!}{\includegraphics{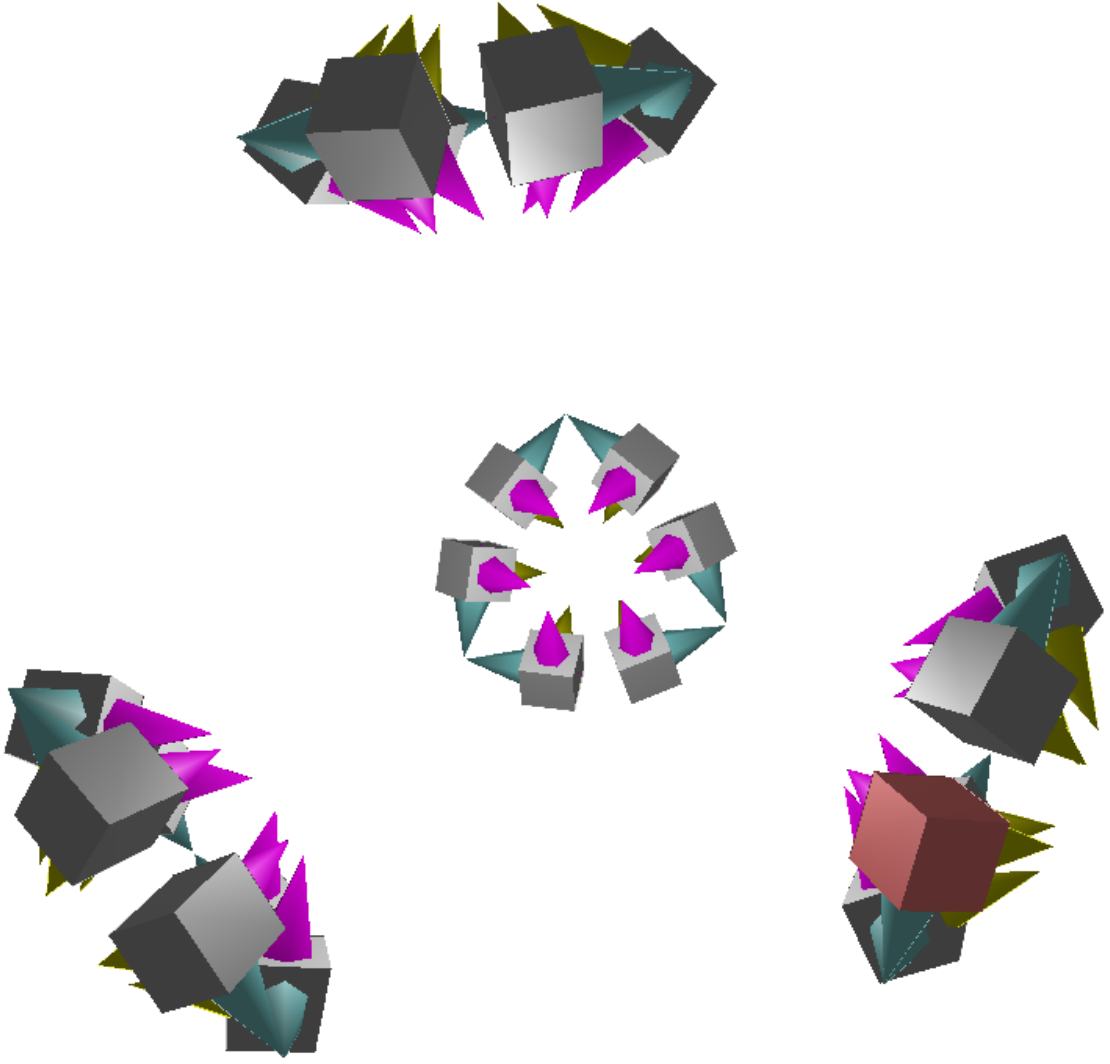}}
  \caption{Left: Diamond cell in Space Group Visualizer. 
           Right: 24 general elements in 3D showing diamond point symmetry of one vertex. 
  \label{fg:Dcell}}
\end{center}
\end{figure}
see Fig. \ref{fg:Dcell} (left), 
or enlarged in Fig. \ref{fg:Dcell} (right). 
In pure diamond one carbon atom is located at the center of this cluster (plus one at $1/4$ distance away along a cubic space diagonal).

The inclusion of integer lattice translations (from fcc vertex to fcc vertex) can lead to new planes of reflection, 
\begin{figure}
\begin{center}
  \resizebox{0.54\textwidth}{!}{\includegraphics{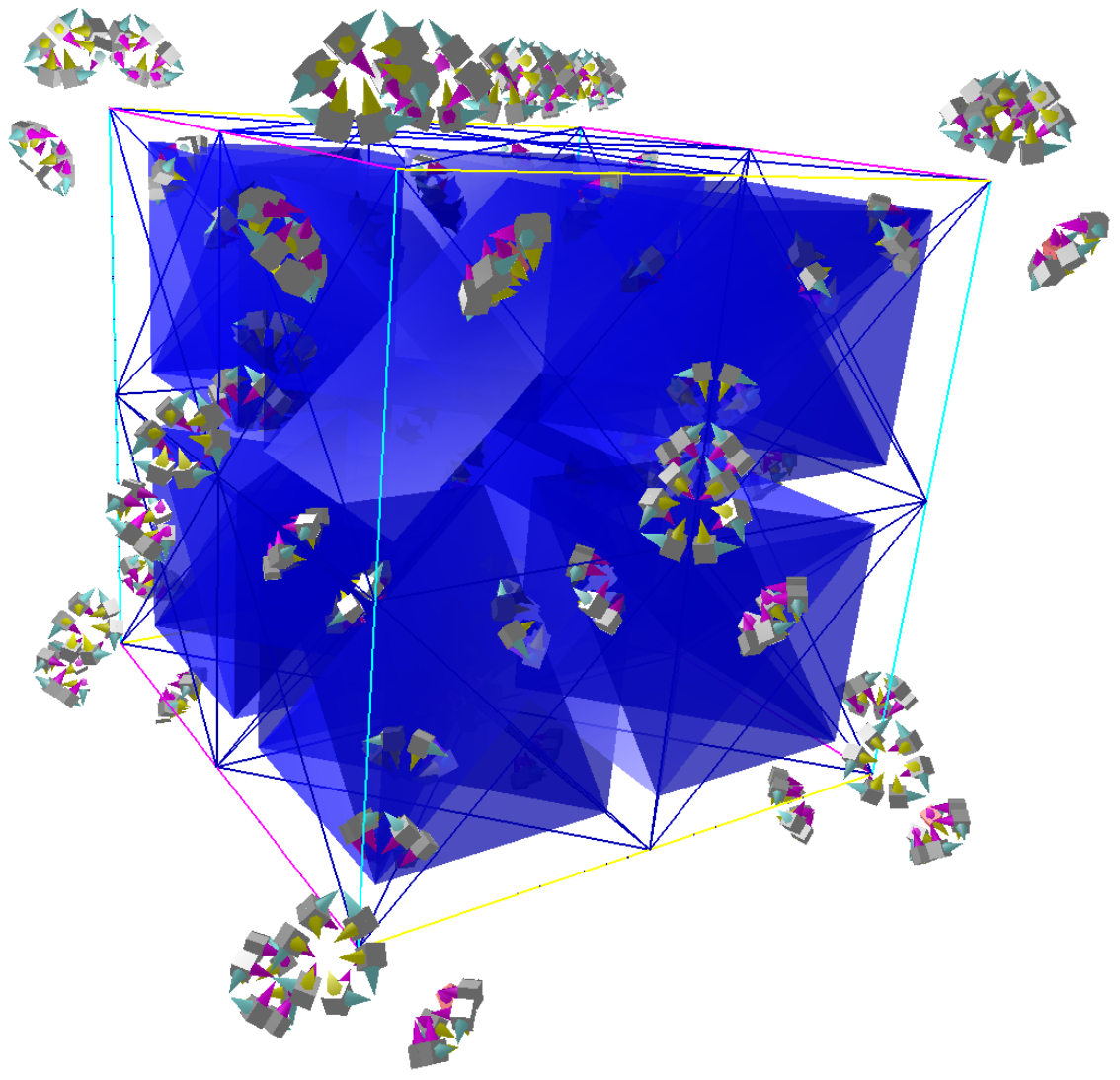}}
  \resizebox{0.44\textwidth}{!}{\includegraphics{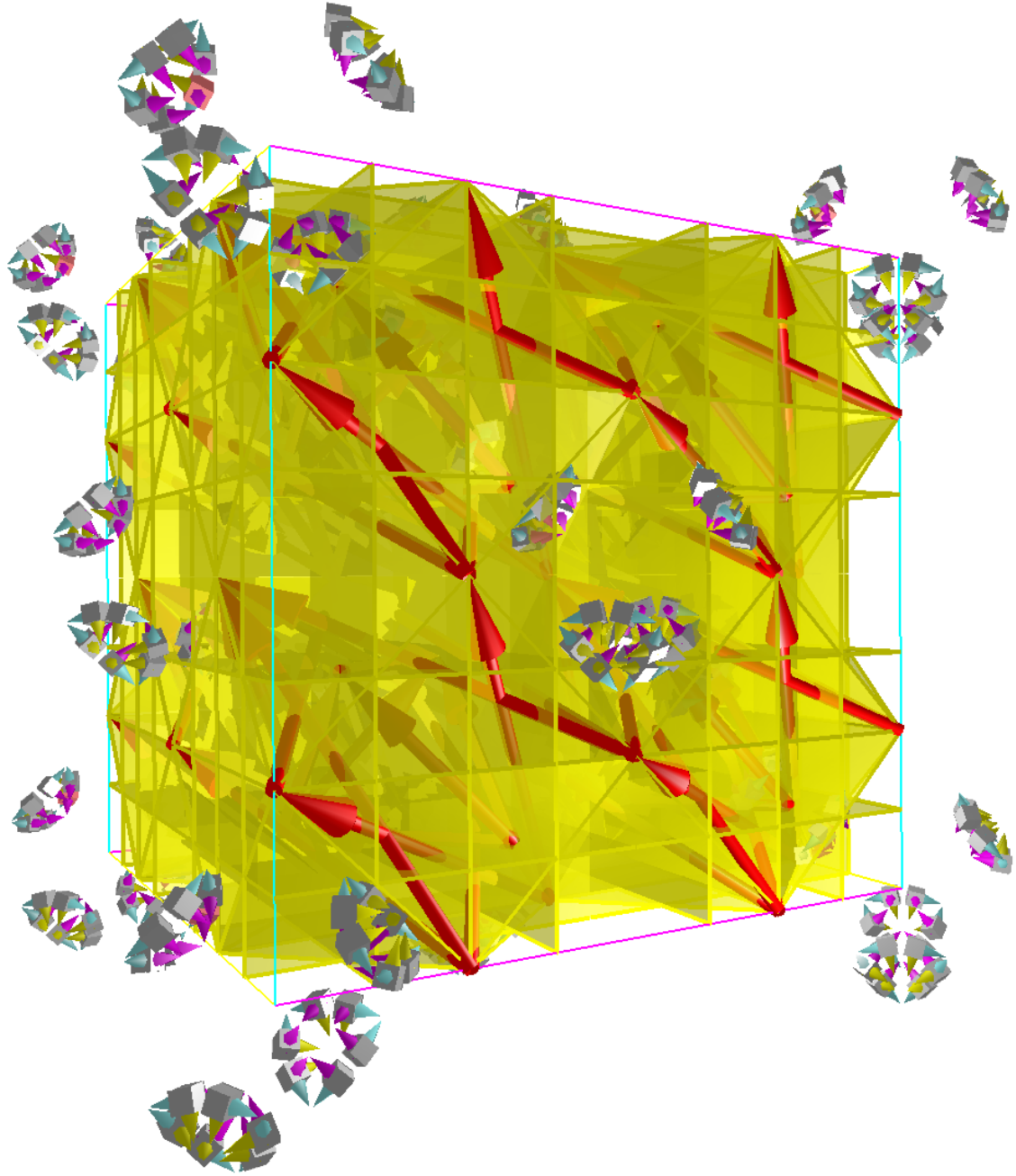}}
  \caption{Left: Diamond reflection planes.
           Right: All glide planes of a diamond cell.
  \label{fg:Refs}}
\end{center}
\end{figure}
see Fig. \ref{fg:Refs} (left). 
The combination of a plane of reflection with a lattice translation not perpendicular to the plane leads to a combined glide reflection, 
see Fig. \ref{fg:Refs} (right), 
where (red) vectors indicate the parallel glide motion. The perpendicular translation component displaces the reflection plane in normal direction, and the parallel translation component creates a glide motion parallel to the plane. Pairs of characteristic diamond glides are 
\begin{figure}
\begin{center}
  \resizebox{0.49\textwidth}{!}{\includegraphics{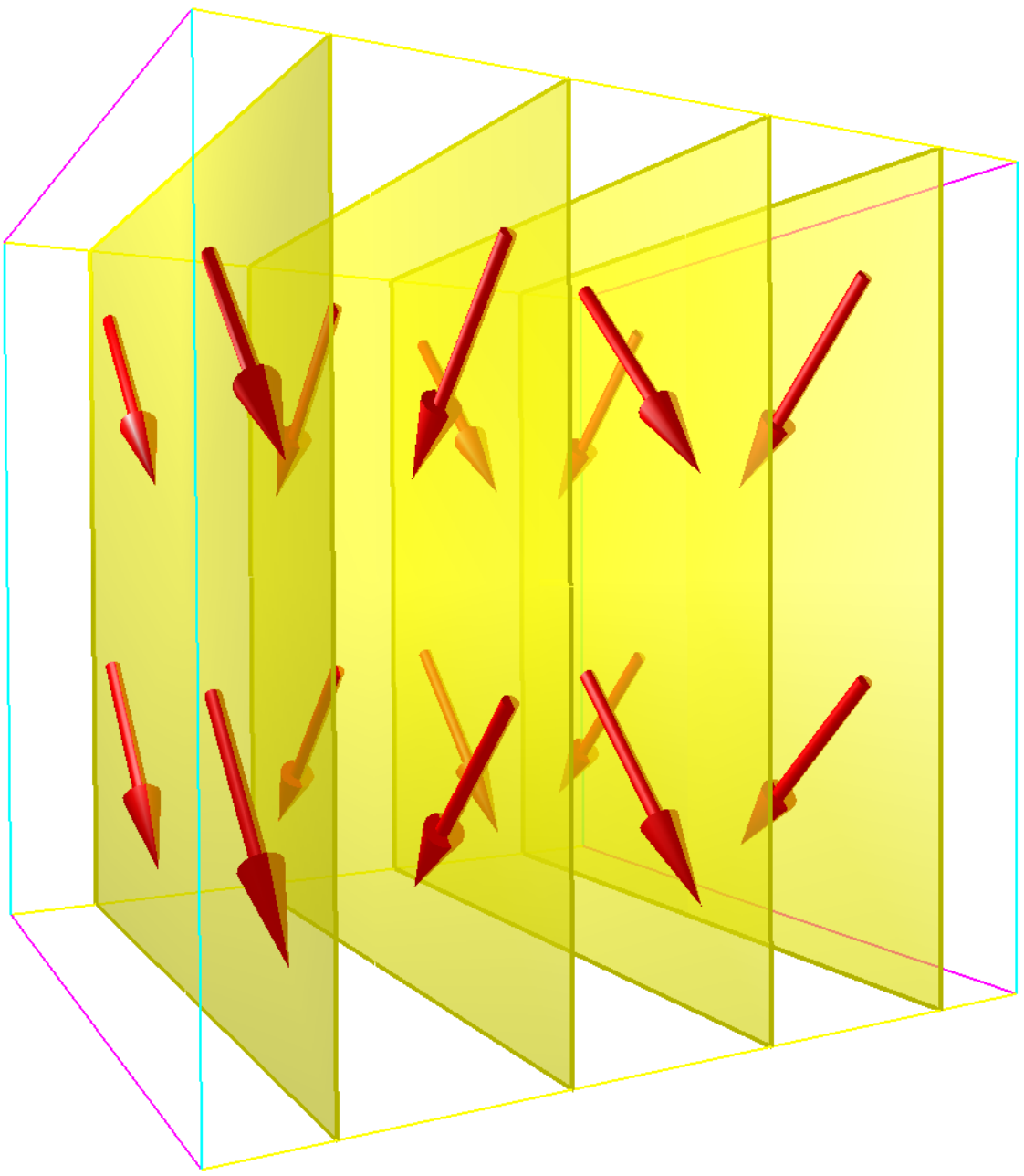}}
  \resizebox{0.49\textwidth}{!}{\includegraphics{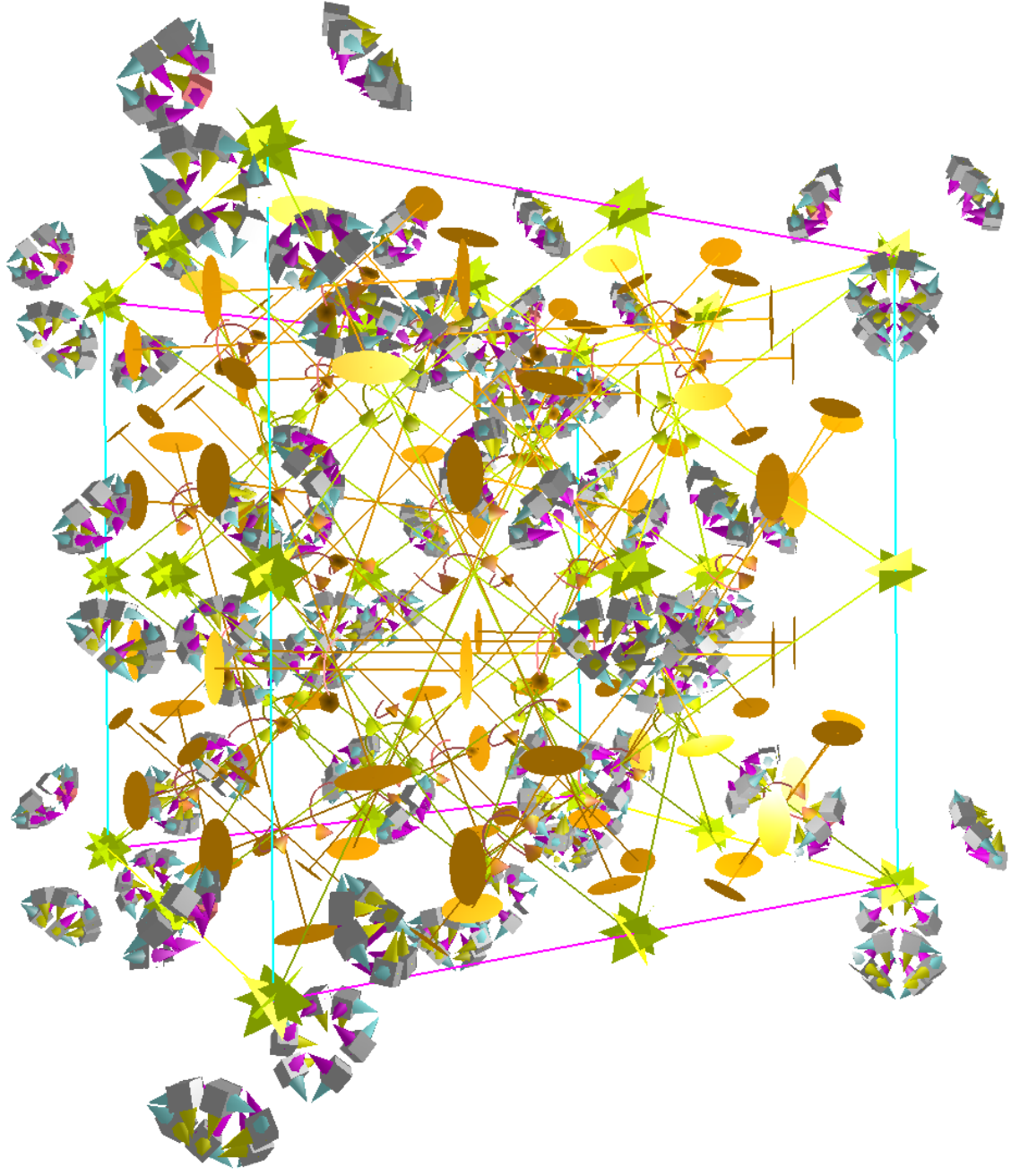}}
  \caption{Left: Pairs of diamond glide planes. 
           Right: Symmetry rotation axis. Angles indicated by colors and arc segments.
  \label{fg:DiamGlides2}}
\end{center}
\end{figure}
shown in Fig. \ref{fg:DiamGlides2} (left). 

A sequence of two reflections at two planes results in a rotation, 
see Fig. \ref{fg:DiamGlides2} (right). 
This rotation has the intersection line of the two planes as its axis and twice the (dihedral) angle between the two planes is the resulting rotation angle. All the rotation axis 
seen in Fig. \ref{fg:DiamGlides2} (right) 
are lines of intersection of reflection planes of 
Fig.  \ref{fg:Refs} (left). 
A lattice translation perpendicular to the rotation axis after a rotation, effectively creates another rotation also already contained in 
Fig. \ref{fg:DiamGlides2} (right). 
But if we perform a translation not normal to the rotation axis, with a translation component parallel to the rotation axis, we get a new transformation, a so-called screw. So a screw is a rotation followed by a translation along the screw axis, resulting in a directed helical motion around the screw axis,
\begin{figure}
\begin{center}
  \resizebox{0.49\textwidth}{!}{\includegraphics{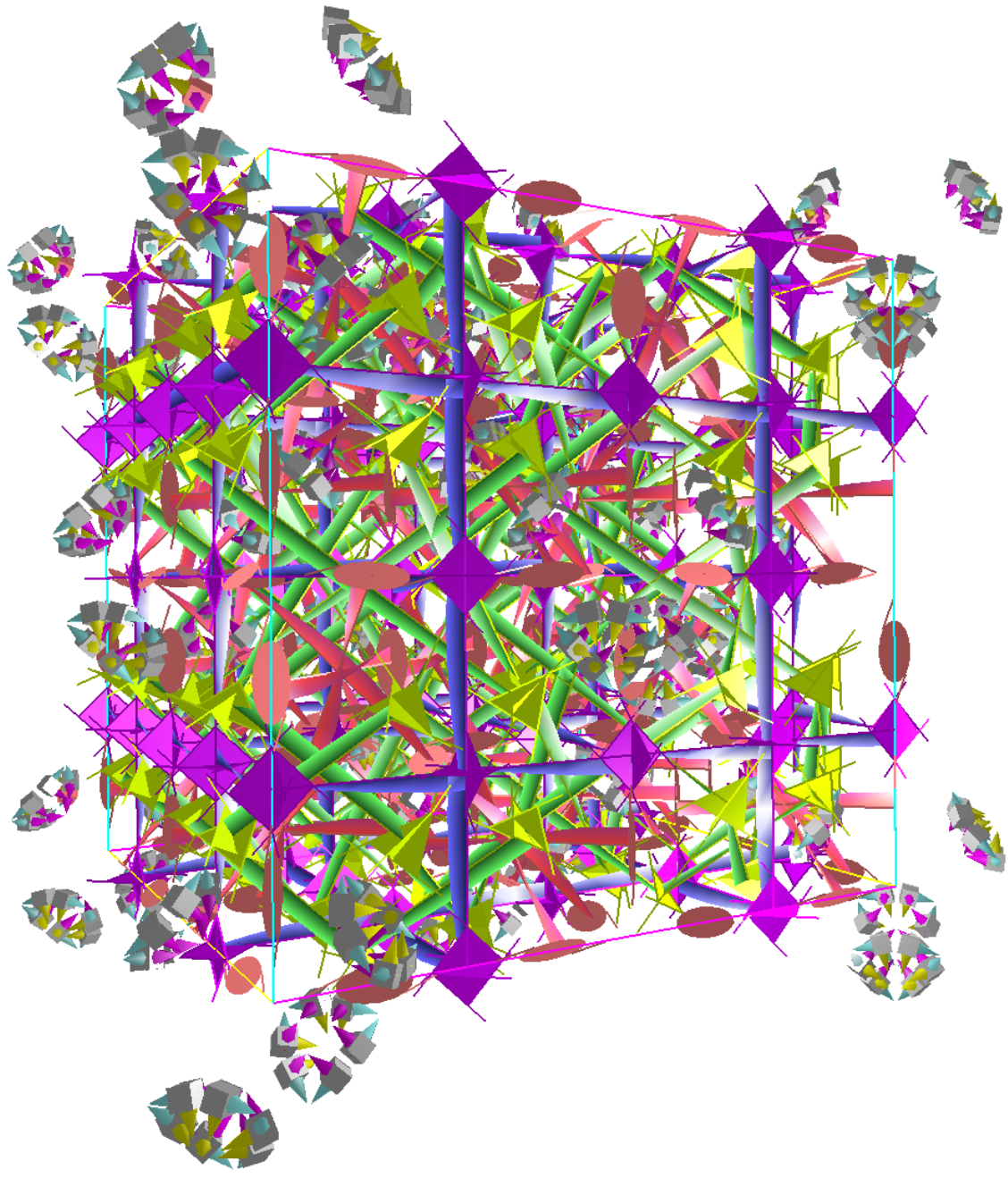}}
  \resizebox{0.49\textwidth}{!}{\includegraphics{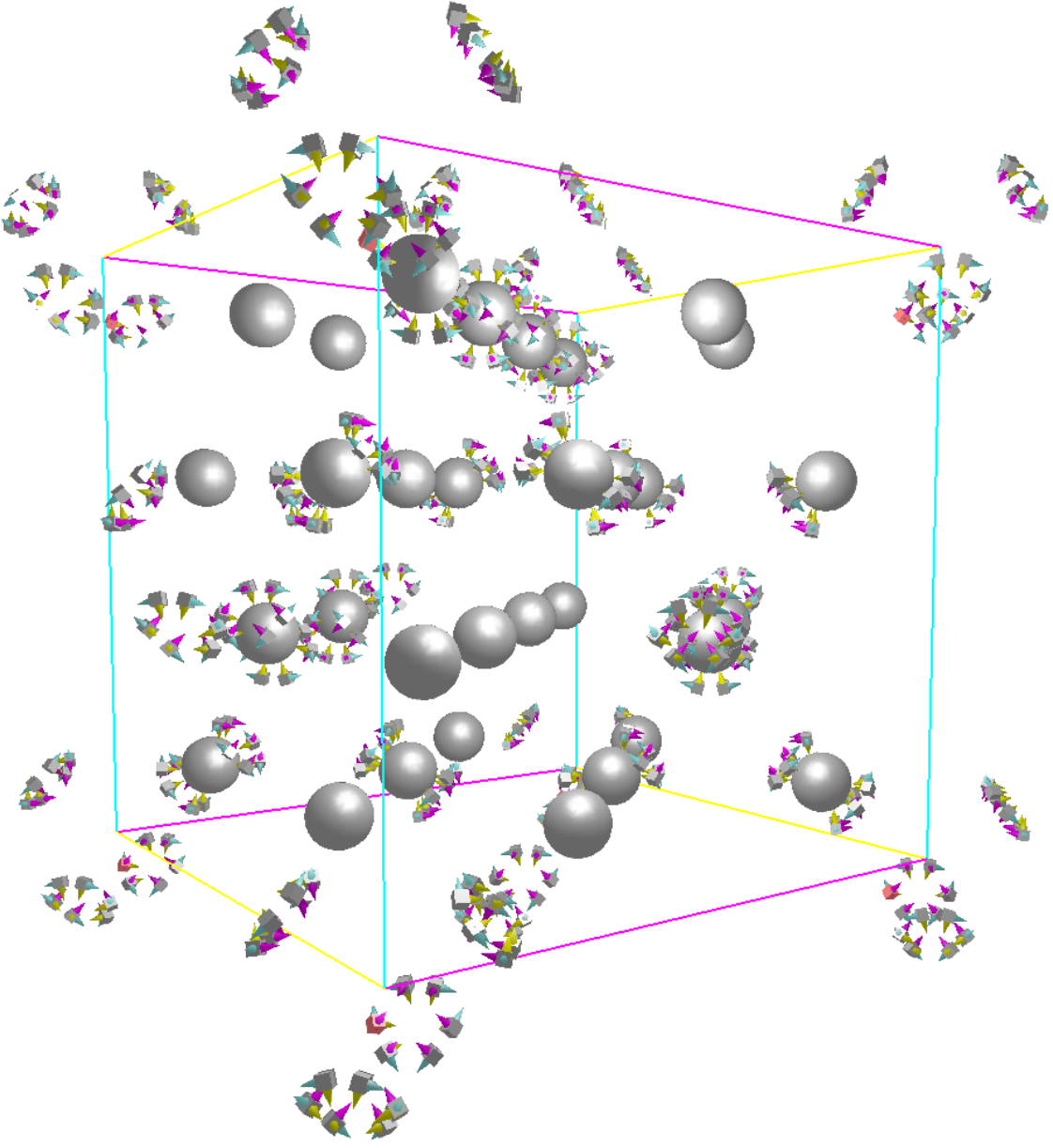}}
  \caption{Left: All screw symmetry axis of diamond.  
           Right: All centers of inversion symmetry of diamond.
  \label{fg:Screws}}
\end{center}
\end{figure}
see Fig. \ref{fg:Screws} (left). 

Combining an inversion with a subsequent lattice translation yields a new center of inversion, 
see Fig. \ref{fg:Screws} (right). 
The combination of an inversion with a rotation leads to a rotoinversion. Characteristic for the diamond lattice are the $90^{\circ}$ rotoinversions depicted in
\begin{figure}
\begin{center}
  \resizebox{0.49\textwidth}{!}{\includegraphics{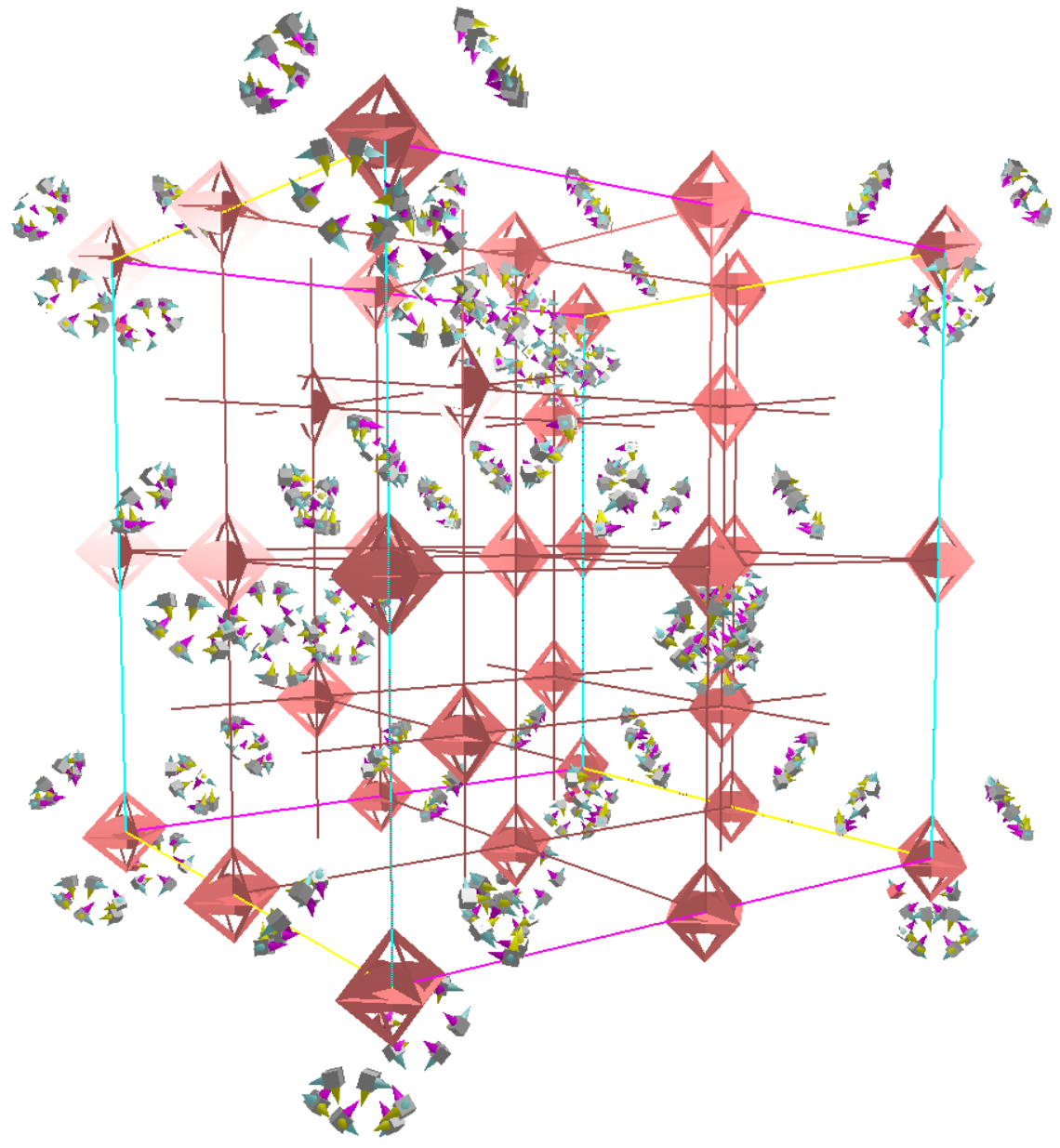}}
  \resizebox{0.49\textwidth}{!}{\includegraphics{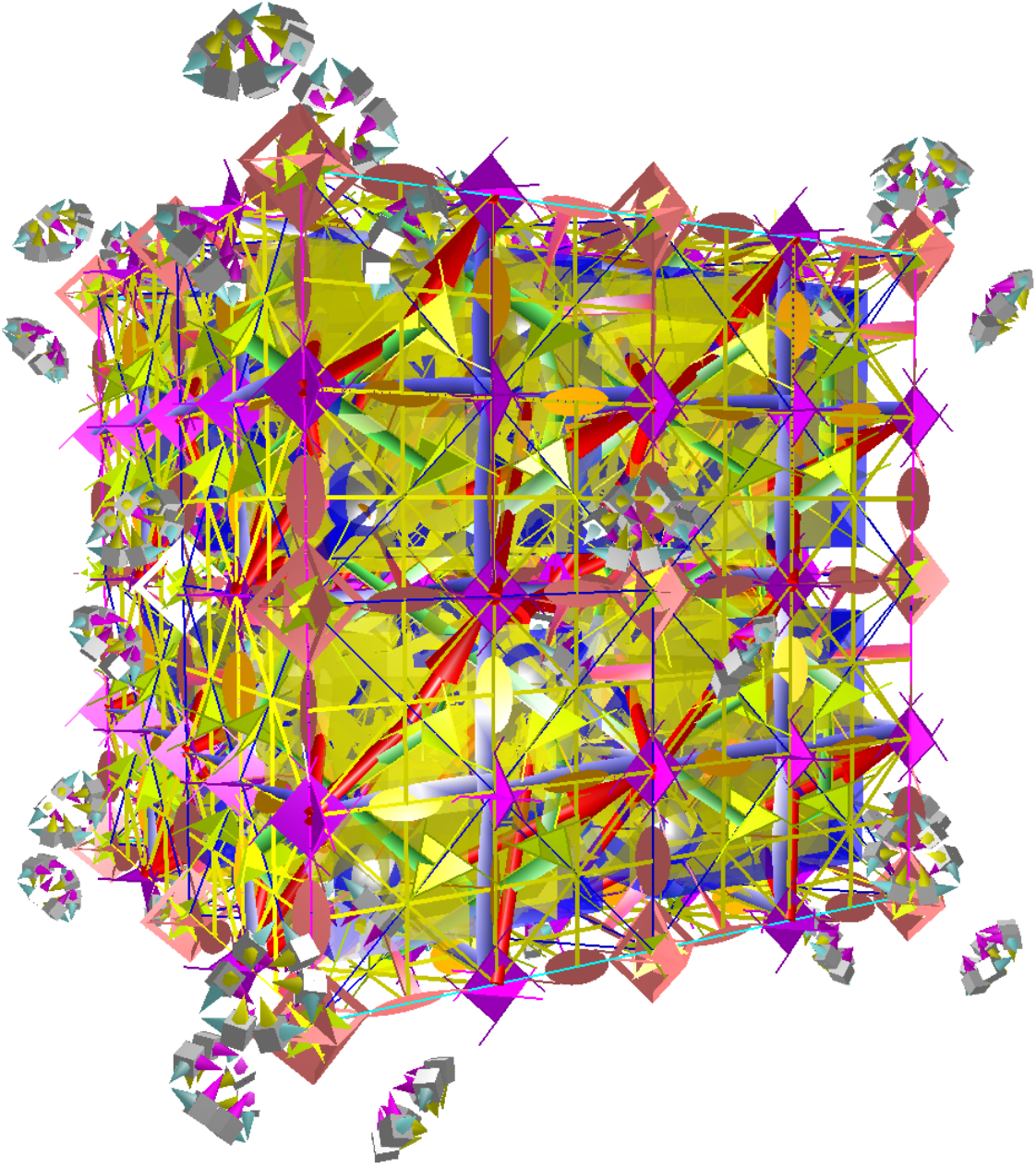}}
  \caption{Left: All centers of inversion symmetry of diamond. 
           Right: Diamond cell in Space Group Visualizer.   
  \label{fg:Rotinvs}}
\end{center}
\end{figure}
Fig. \ref{fg:Rotinvs} (left). 
The total graphical depiction of these symmetries in 
Fig. \ref{fg:Rotinvs}  (right)
gives an idea of the intricate complexity of the symmetries possessed by the diamond lattice. The 
International Tables of Crystallography, Vol. A \cite{TH:ITC}, abbreviated ITA,
depict the symmetries of diamond by showing a quarter of an orthographic 2D projection of a side of a cubic cell. The SGV allows to open an extra window with the ITA online and navigate synchronously in both, 
\begin{figure}
\begin{center}
  \resizebox{0.49\textwidth}{!}{\includegraphics{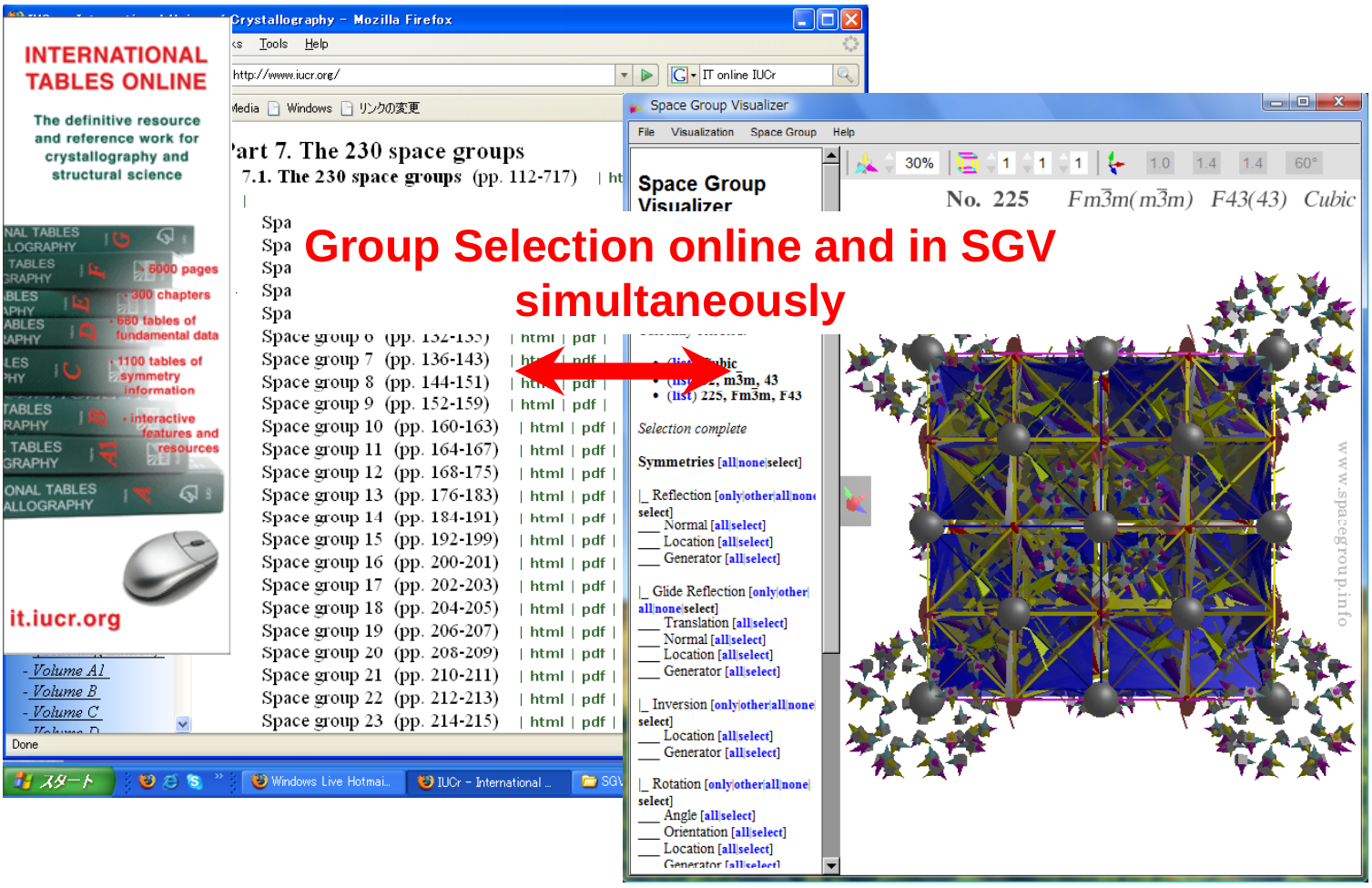}}
  \hspace{5mm}
  \resizebox{0.34\textwidth}{!}{\includegraphics{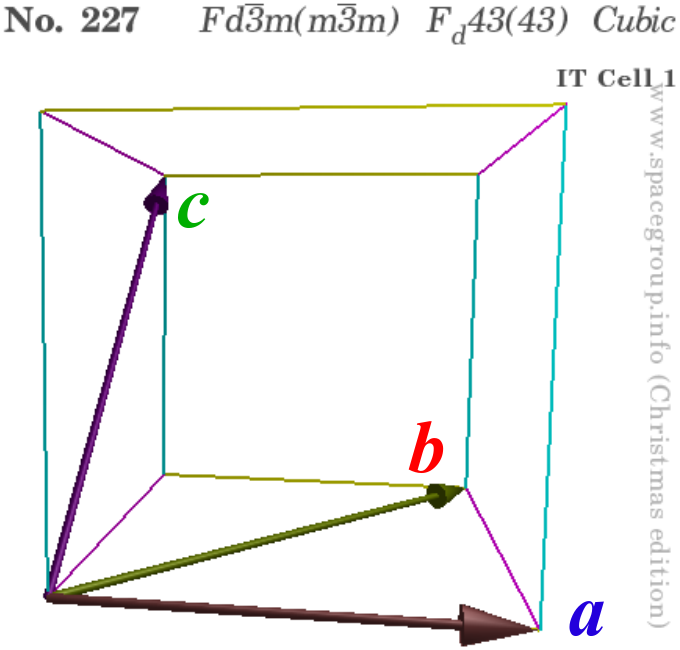}}
  \caption{Left: Space group selection in the online Int. Tables of Crystallography (left half) and the SGV (right half). 
           Right: Three characteristic vectors of diamond cell.
  \label{fg:ITA-SGV}}
\end{center}
\end{figure}
see Fig. \ref{fg:ITA-SGV} (left).

\appendix 
\section{Clifford geometric algebra description of space groups}
  \label{ap:GAspacegr}
  \setcounter{equation}{0}
  \renewcommand{\theequation}{A\arabic{equation}}
  
\subsection{Cartan-Dieudonn\'{e} and geometric algebra \label{sc:Cartan}}

\begin{figure}%[t]
\begin{center}
  \resizebox{0.27\textwidth}{!}{\includegraphics{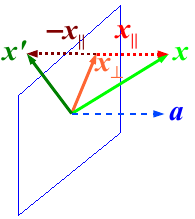}}
  \hspace{2mm}
  \resizebox{0.27\textwidth}{!}{\includegraphics{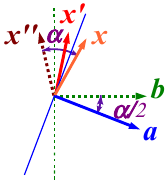}}
  \hspace{2mm}
  \scalebox{0.24}{\includegraphics[width=\textwidth]{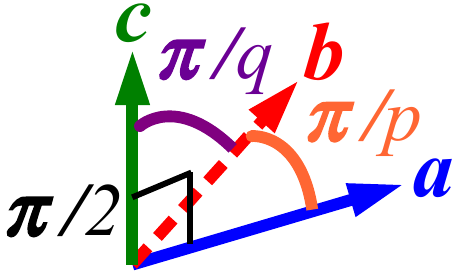}}
  \caption{Left: Reflection at a hyperplane normal to $\gvec{a}$. 
           Center: Rotation generated by two successive reflections 
           at hyperplanes normal
           to $\gvec{a},\gvec{b}$ by twice the 
           angle $\angle(\gvec{a},\gvec{b})$.
           Right: Angular relations of pairs of geometric cell vectors of 
           $\gvec{a},\gvec{b},\gvec{c}$: 
           $\angle(\gvec{a},\gvec{b})=\pi/p$,
           $\angle(\gvec{b},\gvec{c})=\pi/q$,
           $\angle(\gvec{a},\gvec{c})=\pi/2$, $p,q\in \{1,2,3,4,6\}$.
  \label{fg:refrot}}
\end{center}
\end{figure}

Clifford's associative geometric product~\cite{WC:AppGExtAlg} of two vectors simply adds 
the (symmetric) inner product to the (anti-symmetric) outer product of Grassmann
\begin{equation}
  \gvec{a}\gvec{b} = \gvec{a}\cdot\gvec{b} + \gvec{a}\wedge\gvec{b}\,.
  \label{eq:gp}
\end{equation}
The mathematical meaning of the left and right side of \eqref{eq:gp} is clear from applying the geometric product to the $n$ orthonormal basis vectors $\{\gvec{e}_1, \ldots , \gvec{e}_n\}$ of the underlying vector space $\mathbb{R}^{p,q}, n=p+q$. We thus have
\begin{align}
  \gvec{e}_k\gvec{e}_k &= 
  \gvec{e}_k\cdot\gvec{e}_k=+1, 
  \quad \gvec{e}_k\wedge\gvec{e}_k=0,\quad 1\leq k \leq p, \\
  \gvec{e}_k\gvec{e}_k &=  
  \gvec{e}_k\cdot\gvec{e}_k=-1,
  \quad \gvec{e}_k\wedge\gvec{e}_k=0, 
  \quad p+1\leq k \leq n, \\
  \gvec{e}_k\gvec{e}_l\, &= -\gvec{e}_l\gvec{e}_k=\gvec{e}_k\wedge\gvec{e}_l,
  \quad \gvec{e}_k\cdot\gvec{e}_l=0,
  \quad l\neq k, \,\, 1\leq k,l \leq n.
\end{align}
Under this product parallel vectors commute and perpendicular vectors anti-commute
\begin{equation}
  \gvec{a}\gvec{x}_{\parallel} = \gvec{x}_{\parallel}\gvec{a}\,,
  \quad \quad
  \gvec{a}\gvec{x}_{\perp} = - \gvec{x}_{\perp}\gvec{a}\,.
\end{equation}
This allows to write the \textit{reflection} of a vector $\gvec{x}$ \textit{at a hyperplane}
through the origin with normal $\gvec{a}$ as (see left side of Fig. \ref{fg:refrot})
\begin{equation}
  \label{eq:ref}
  \gvec{x}^{\,\,\prime}= -\,\gvec{a}^{\,\, -1} \gvec{x}\,\gvec{a}\, ,
  \quad \quad
   \gvec{a}^{\,\, -1} = \frac{\gvec{a}}{\gvec{a}^{\,\, 2}} \, .
\end{equation}
The composition of two reflections at hyperplanes, whose normal vectors
$\gvec{a}, \gvec{b}$
subtend the angle $\alpha/2$, yields a rotation around the intersection of
the two hyperplanes (see center of Fig. \ref{fg:refrot}) by $\alpha$
\begin{equation}
  \label{eq:rot}
  \gvec{x}^{\,\,\prime\prime}
  = (\gvec{a}\gvec{b})^{-1} \gvec{x}\,\gvec{a}\gvec{b} \, ,
  \quad \quad
  (\gvec{a}\gvec{b})^{-1} = \,\gvec{b}^{\,\, -1}\,\gvec{a}^{\,\, -1} \, .
\end{equation}

Continuing with a third reflection at a hyperplane with normal
$\gvec{c}$ according to the Cartan--Dieudonn\'{e} 
theorem 
yields rotary reflections (equivalent to rotary inversions
with angle $\alpha - \pi$) and inversions
\begin{equation}
  \label{eq:rotoinv}
 \gvec{x}^{\,\,\prime}
  = -\,(\gvec{a}\gvec{b}\gvec{c})^{-1} \gvec{x}\,\gvec{a}\gvec{b}\gvec{c} \, ,
  \quad \quad
 \gvec{x}^{\,\,\prime\prime}
  = -\,i^{-1} \gvec{x} \,i \, ,
  \quad \quad
  i \doteq \gvec{a}\wedge \gvec{b}\wedge \gvec{c},
\end{equation}
where $\doteq$ means equality up to non-zero scalar factors 
(which cancel out in \eqref{eq:symtrafo}). 
In general the geometric product of $k$ normal vectors 
(the versor $S$) 
corresponds to the
composition of reflections to all symmetry transformations~\cite{HH:CSG} of 
two-dimensional (2D) and 3D crystal cell point groups (also called crystal classes)
\begin{equation}
  \label{eq:symtrafo}
 \gvec{x}^{\,\,\prime}
  = (-1)^k S^{\,-1} \,\gvec{x} \,S.
\end{equation}

\subsection{Two dimensional point groups}

\begin{figure}%[t]
\begin{center}
  \resizebox{0.8\textwidth}{!}{\includegraphics{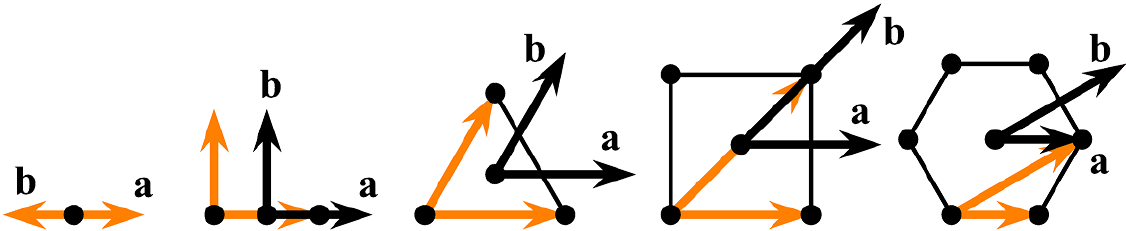}}
  \caption{Regular polygons ($p=1,2,3,4,6$) and point 
    group generating vectors $\gvec{a},\gvec{b}$ subtending 
    angles $\pi/p$ shifted to center. \label{fg:2Dpg}}
\end{center}
\end{figure}

2D point groups~\cite{HH:CSG} are generated by multiplying vectors 
selected~\cite{HP:crystGA, HP:sym3vec, HP:intvis} as in 
Fig. \ref{fg:2Dpg}. The index $p$ denotes these groups as in Table 
\ref{tb:2Dpg}. 
For example the trigonal (square) point group (the symmetry group
of the reg. triangle (square), leaving the center point invariant) is given by multiplying its
two generating vectors $\gvec{a},\gvec{b}$
\begin{align}
 \label{eq:3}
 3 &= \{
 \gvec{a}, \gvec{b}, R=\gvec{a}\gvec{b}, R^2, R^3=-1, 
 \gvec{a}R^2 \}. 
 \\
 \label{eq:4}
 4 &= \{
 \gvec{a}, \gvec{b}, R=\gvec{a}\gvec{b}, R^2, R^3, R^4=-1, 
 \gvec{a}R^2, \gvec{b}R^2 \}.
\end{align}
In \eqref{eq:3} and \eqref{eq:4} the vectors $\gvec{a},\gvec{b}$ represent reflections \eqref{eq:ref} 
at lines 
normal to $\gvec{a},\gvec{b}$ and passing through the center of the reg. triangle (square) of Fig. \ref{fg:2Dpg}.
The rotor $R=\gvec{a}\gvec{b}$ represents as in \eqref{eq:rot} a double reflection at the two lines passing through the center and normal to $\gvec{a}$ and $\gvec{b}$, respectively. 
Because $\angle(\gvec{a},\gvec{b})=60^\circ \, (45^\circ)$, the resulting rotation
is by $2\times 60^\circ = 120^\circ \, (2\times 45^\circ = 90^{\circ})$ around the center. 
The cyclic rotation subgroups are denoted in Table \ref{tb:2Dpg} with bars, e.g. 
\begin{equation}
  \bar{3} = \{R=\gvec{a}\gvec{b}, R^2, R^3=-1 \doteq 1\}, \quad
  \bar{4} = \{R=\gvec{a}\gvec{b}, R^2, R^3, R^4=-1 \doteq 1\},
\end{equation}
containing the three (four) symmetry rotations of the
reg. triangle (square) of Fig. \ref{fg:2Dpg} around its invariant center by 
$120^\circ$, and the multiples $240^\circ$ and $360^\circ$ 
($90^{\circ}, 180^{\circ}, 270^{\circ}$ and $360^{\circ}$).
The vectors  
$\gvec{a}R^2$ in \eqref{eq:3} 
($\gvec{a}R^2, \gvec{b}R^2 $ in \eqref{eq:4})
are the normal directions of the remaining one (two) lines (passing through the center) of reflection symmetry.

\begin{table}%[ht]
\begin{center}
\caption{Geometric and international notation for 2D point groups.\label{tb:2Dpg}}
\begin{tabular}{lllllllllll}
%\hline\noalign{\smallskip}
\topline
Crystal&\multicolumn{2}{c}{Oblique}
&\multicolumn{2}{c}{Rectangular}
&\multicolumn{2}{c}{Trigonal}
&\multicolumn{2}{c}{Square}
&\multicolumn{2}{c}{Hexagonal}
\\
\midline
\rule{0mm}{3.7mm}%
%\noalign{\smallskip}\svhline\noalign{\smallskip}
geometric & $\bar{1}$& $\bar{2}$ & 1& 2 & 3& $\bar{3}$ & 4&$\bar{4}$ & 6& $\bar{6}$
\\
international  &1& 2 &m& mm& 3m& 3& 4m& 4& 6m& 6\\
\bottomline
\end{tabular}
\end{center}
\end{table}

\subsection{Three dimensional point groups}

The selection of three characteristic vectors $\gvec{a},\gvec{b},\gvec{c}$ (see Fig. \ref{fg:ITA-SGV} (right) for diamond)
from each crystal cell is sufficient~\cite{HH:CSG, HP:crystGA, HP:sym3vec, HP:intvis} for generating all 3D point groups. 

For the purpose of point groups keeping a single cell 
\textit{as a whole invariant}, the vectors
$\gvec{a},\gvec{b},\gvec{c}$ have always to be attached to the invariant 
cell center. These three vectors are normals of characteristic
planes passing through the cell center. 
The plane reflections which the vectors represent
and their combinations as in \eqref{eq:ref} to \eqref{eq:symtrafo} constitute all point symmetries of the 3D crystal cells. 
The point symmetry transformations are applied to every vertex of a cell
and keep the cell as a whole invariant, transforming each vertex into
another vertex. 

Using $\angle(\gvec{a},\gvec{b})$ and $\angle(\gvec{b},\gvec{c})$
(right side of Fig. \ref{fg:refrot}) 
we can denote all 32 3D point groups as in Table \ref{tb:3Dpg} ($pq=43$ for diamond).
Again the overbar notation, e.g. $\bar{p}$, means that
the two vectors concerned are only to be used in their fixed 
rotor combination, e.g. $\gvec{a}\gvec{b}$. If the closed overbar
extends over both indexes $\overline{pq}$ all three vectors
are only to be used in the fixed rotoinversion (alias rotary reflection)
combination $\gvec{a}\gvec{b}\gvec{c}$ of \eqref{eq:rotoinv}.

Note that the notation of Table \ref{tb:3Dpg}
is fully isomorphic to the notation used in Table 2 of \cite{CM:GRDG},
with $p, q$ in the very same roles.

\begin{table}%[t]
\begin{center}
\caption{\label{tb:3Dpg}
Geometric 3D point group symbols~\cite{HH:CSG} and generators with
$\gvec{a},\gvec{b},\gvec{c}$: 
$\angle(\gvec{a},\gvec{b})=\pi/p$,
$\angle(\gvec{b},\gvec{c})=\pi/q$,
$\angle(\gvec{a},\gvec{c})=\pi/2$, $p,q\in \{1,2,3,4,6\}$.}
\begin{tabular}{lccccccccc}
\topline
\rule{0mm}{4mm}%
Symbol
& $p=1$ & $p$ & $\bar{p}=\bar{1}$ &  $\bar{p}$ & $pq$ 
& $\bar{p}q$ & $p\bar{q}$ & $\bar{p}\bar{q}$ & $\overline{pq}$
\\
\midline
Generators 
& $\, \gvec{a} \,$
& $\, \gvec{a}, \gvec{b} \,$
& $\, 1 \, $
& $\,\gvec{a}\gvec{b} \,$ 
& $\,\gvec{a}, \gvec{b}, \gvec{c} \,$
& $\,\gvec{a}\gvec{b}, \gvec{c} \,$ 
& $\,\gvec{a}, \gvec{b}\gvec{c} \,$
& $\,\gvec{a}\gvec{b}, \gvec{b}\gvec{c} \,$ 
& $\,\gvec{a}\gvec{b}\gvec{c}$
\\
\bottomline
\end{tabular}
\end{center}
\end{table}

\subsection{Space groups \label{sc:SG}}

The smooth composition with translations is best done in the conformal
model~\cite{SL:Diss, LVA:MoebRn, HL:IAGR} of Euclidean space (in the GA of $\mathbb{R}^{4,1}$), 
which adds two null-vector dimensions for 
the origin $\gvec{e}_0$ and infinity $\gvec{e}_{\infty}$
\begin{equation}
  X = \gvec{x} + \frac{1}{2}\gvec{x}^2\gvec{e}_{\infty}+\gvec{e}_0, 
  \quad
  \gvec{e}_0^2=\gvec{e}_{\infty}^2=X^2=0,
  \quad
  X\cdot \gvec{e}_{\infty}=-1.
\end{equation}
The inner product of two conformal points gives their Euclidean distance and 
therefore a plane $m$ equidistant from two points $A,B$ as
\begin{equation}
  X\cdot A = -\frac{1}{2}(\gvec{x}-\gvec{a})^2 \,\,
  \Rightarrow \,\,X\cdot (A-B)=0,
  \quad
  m=A-B \propto \gvec{n}-d\,\gvec{e}_{\infty},
\end{equation}
where $\gvec{n}$ is a unit normal to the plane and $d$ its signed scalar distance 
from the origin. Reflecting at two parallel planes $m,m^{\prime}$ with 
distance $\gvec{t}/2$ we get the \textit{transla}tion opera\textit{tor} 
(by $\gvec{t}\,$)
\begin{equation}
  X^{\prime} = m^{\prime}m\,X\,mm^{\prime} = T_{\gvec{t}}^{-1} X T_{\gvec{t}},
  \quad T_{\gvec{t}}=1+\frac{1}{2}\gvec{t}\gvec{e}_{\infty}.
\end{equation}
Reflection at two non-parallel planes $m,m^{\prime}$ yields the rotation around
the $m,m^{\prime}$-intersection by twice the angle subtended by $m,m^{\prime}$.

Group theoretically the conformal group $C(3)$ is isomorphic to $O(4,1)$ and the
Euclidean group $E(3)$ is the subgroup of $O(4,1)$ leaving infinity 
$\gvec{e}_{\infty}$ invariant. 
Now general translations and rotations are represented by geometric products
of invertible vectors (called Clifford monomials, Lipschitz elements, or \textit{versors}). 

Applying these techniques one can compactly tabulate geometric space group 
symbols and generators~\cite{HH:CSG}. Diamond has space group $F_d43$ with generators 
$\{\gvec{a}T_{\gvec{c}/4}, \gvec{b}, \gvec{c},  T_{\gvec{a}}, T_{\gvec{b}/2}, T_{\gvec{c}/2} \}$ 
%$\{\gvec{a}T_{\gvec{c}}^{1/4}, \gvec{b}, \gvec{c},  T_{\gvec{a}}, T_{\gvec{b}}^{1/2}, T_{\gvec{c}}^{1/2} \}$ 
(SGV correction to~\cite{HH:CSG}!).

% ------------------------------------------------------------------------

\subsection*{Acknowledgment}
I wish to thank God for his wonderful creation with the words of H. Grassmann:
%\textit{You answer us with awesome deeds of righteousness,
%       O God our Savior,
%       the hope of all the ends of the earth
%       and of the farthest seas ...}~(Ps. 65:5)\cite{Psalm65:5}.
%%%
\textit{Ich
glaube also den in der Bibel geoffenbarten Wahrheiten
nicht darum, weil sie in der Bibel stehen, sondern weil
ich ihre seligmachende Kraft, ihre ewige, g\oe{}ttliche
Wahrheit in meinem Bewu\ss{}tsein erfahren habe.} \cite{HG:AbfallvG}
(English: I therefore believe the truths revealed in the Bible, 
not because they are written in the Bible, 
but because I have experienced in my own conscience
their power of blessing, their eternal, divine truth.)
%%%
I thank 
my family for their loving support, 
O. Giering, D. Hestenes, C. Perwass, M. Aroyo, D. Litvin, and H.-J. Petsche.

% ------------------------------------------------------------------------
\subsection*{Dr. Eckhard Hitzer}
\textit{Educational background:} 
Ph. D. in Theoretical Physics.
\\
\textit{Current position:} 
Part Time Lecturer at the University of Fukui.
\\
\textit{Main research interest:} 
Theory and applications of geometric calculus, crystallography, visualization, neural computation.
\\
\textit{Selected publications:} 
Real Clifford Algebra $Cl(n,0), n = 2,3(\mathrm{mod} 4)$ Wavelet Transform (2009);
Geometric Roots of $-1$ in Clifford Algebras $Cl(p,q)$ with $p+q <= 4$ (2009), with R. Ab\l{}amowicz;
Interactive 3D Space Group Visualization with CLUCalc and the Clifford Geometric Algebra Description of Space Groups (2008), with C. Perwass.
\\
\textit{Address:} Department of Applied Physics, University of Fukui, 3-9-1 Bunkyo, 910-8507 Fukui, Japan.
\\
\textit{E-mail:} \email{hitzer@mech.u-fukui.ac.jp}
% ------------------------------------------------------------------------
\end{document}